\documentclass[aps,pra,superscriptaddress,amsmath,amssymb,twocolumn,amsfonts,floatfix, longbibliography]{revtex4-2}
\usepackage{graphicx}
\usepackage{epstopdf}
\usepackage[T1]{fontenc}
\usepackage[latin9]{inputenc}
\usepackage{amsbsy}
\usepackage{gensymb}

\usepackage[dvipsnames]{xcolor}
\definecolor{deepfuchsia}{rgb}{0.76, 0.33, 0.76}
\definecolor{electricpurple}{rgb}{0.75, 0.0, 1.0}
\usepackage[colorlinks=true,linktoc=page,linkcolor=blue,citecolor=blue,urlcolor=blue]{hyperref}
\usepackage{orcidlink}
\usepackage[normalem]{ulem}

\usepackage{titlesec} 

\titleformat{\section}
  {\raggedright\bfseries\large} 
  {\thesection}{1em}{}

\titleformat{\subsection}
  {\raggedright\bfseries\large} 
  {\thesubsection}{1em}{}

\titlespacing*{\section}
  {0pt} 
  {5pt} 
  {0pt} 

\titlespacing*{\subsection}
  {0pt} 
  {3pt} 
  {0pt} 

\newcommand{\beq}{\begin{equation}}
\newcommand{\eeq}{\end{equation}}
\newcommand{\bea}{\begin{eqnarray}}
\newcommand{\eea}{\end{eqnarray}}

\newcommand{\figref}[1]{Fig.~\ref{#1}}
\newcommand{\refcite}[1]{Ref.~\onlinecite{#1}}

\renewcommand{\vec}[1]{\boldsymbol{#1}}

\renewcommand\thesection{\arabic{section}}
\renewcommand{\thesubsection}{\alph{subsection}}

\begin{document}
\title{Ising superconductivity in bulk layered non-centrosymmetric 4$H$-NbSe$_2$}

\author{Chandan Patra}\thanks{These authors contributed equally to this work}
\affiliation{Department of Physics, Indian Institute of Science Education and Research Bhopal, Bhopal, 462066, India}

\author{Tarushi Agarwal}\thanks{These authors contributed equally to this work}
\affiliation{Department of Physics, Indian Institute of Science Education and Research Bhopal, Bhopal, 462066, India}

\author{Rahul Verma}\thanks{These authors contributed equally to this work}
\affiliation{Department of Condensed Matter Physics and Materials Science, Tata Institute of Fundamental Research, Mumbai 400005, India}

\author{Poulami Manna}
\affiliation{Department of Physics, Indian Institute of Science Education and Research Bhopal, Bhopal, 462066, India}

\author{Shashank Srivastava}
\affiliation{Department of Physics, Indian Institute of Science Education and Research Bhopal, Bhopal, 462066, India}

\author{Ravi Shankar Singh}
\affiliation{Department of Physics, Indian Institute of Science Education and Research Bhopal, Bhopal, 462066, India}

\author{Mathias~S.~Scheurer}
\affiliation{Institute for Theoretical Physics III, University of Stuttgart, 70550 Stuttgart, Germany}

\author{Bahadur Singh}\
\email[]{bahadur.singh@tifr.res.in}
\affiliation{Department of Condensed Matter Physics and Materials Science, Tata Institute of Fundamental Research, Mumbai 400005, India}

\author{{Ravi Prakash Singh}\,\orcidlink{0000-0003-2548-231X}}
\email[]{rpsingh@iiserb.ac.in}
\affiliation{Department of Physics, Indian Institute of Science Education and Research Bhopal, Bhopal, 462066, India}

\begin{abstract}
\begin{flushleft}
\end{flushleft}
{
Transition metal dichalcogenides exhibit multiple polymorphs that enable the exploration of diverse quantum states, including valley-selective spin polarization, the valley Hall effect, Ising superconductivity, and nontrivial topology. Monolayer 2$H$-NbSe$_2$ is a promising candidate for realizing Ising superconductivity due to its spin-split, out-of-plane spin-polarized states arising from inversion symmetry breaking and strong spin-orbit coupling. In contrast, bulk 2$H$-NbSe$_2$ retains inversion symmetry and lacks spin splitting, limiting its suitability for hosting Ising superconductivity. Here, we report the growth of high-quality single crystals of the acentric bulk superconducting polymorph, 4$H$-NbSe$_2$, which intrinsically breaks the inversion symmetry and supports valley-selective spin-polarized states. Magnetization and resistivity measurements reveal anisotropic superconductivity, with the in-plane upper critical field exceeding the Pauli limit, while out-of-plane fields suppress superconductivity more rapidly, before reaching the Pauli limit, which strongly suggests the presence of Ising pairing. First-principles calculations and symmetry analysis confirm significant valley-selective spin splitting with out-of-plane spin polarization, further supporting the emergence of Ising superconductivity in 4$H$-NbSe$_2$. These results establish 4$H$-NbSe$_2$ as a robust bulk platform to investigate Ising superconductivity and valley-selective phenomena in transition-metal dichalcogenides.
}
\end{abstract}

\maketitle
 \section{Introduction}
 Transition metal dichalcogenides (TMDs) exhibit a diverse range of electronic phases, including semiconducting, metallic, semimetallic, and superconducting, arising from the partally filled $d$-bands of transition metals and their intrinsically low structural dimensionality. These materials provide an ideal platform for studying exotic quantum phenomena, such as the interplay between electron spin, layered pseudospin, and momentum, and for advancing applications in spintronics, optoelectronics, and quantum technologies~\cite{semi_conducting,unconvention_sc,topology_tmds,spin_psudospin}. Monolayer TMDs, with their fundamental trigonal prismatic arrangement, break the in-plane inversion symmetry by sandwiching layers of transition metal atoms between chalcogen atoms~\cite{inversion_symmetry}. The absence of crystal inversion symmetry, combined with the strong spin-orbit coupling (SOC) of the transition metal atoms, generates an effective Zeeman-type spin-orbit magnetic field~\cite{breaking_inversion_symmetry}. This field causes electron spins to align out of the plane, known as Ising SOC, with opposite signs for opposite momenta at the $K$ and $K'$ points of the hexagonal Brillouin zone~\cite{aling_of_k_k_dash,graphene_symmetry,ising_sc_review,bi2se3_nbse2_hetero}. In the superconducting state, this unique spin-momentum locking leads to Ising superconducting pairing, which has recently been observed in monolayer 1$H$-NbSe$_2$ \cite{ising_nbse2}, 1$H$-TaS$_2$ \cite{TaS2_ising}, and gated MoS$_2$ \cite{MoS2_ising}. The entanglement of spin and momentum in this superconducting pairing inhibits spin pair-breaking, leading to remarkably high in-plane upper critical fields that exceed the Pauli limit.

\begin{figure*}[t!]
\includegraphics[width=0.98\textwidth]{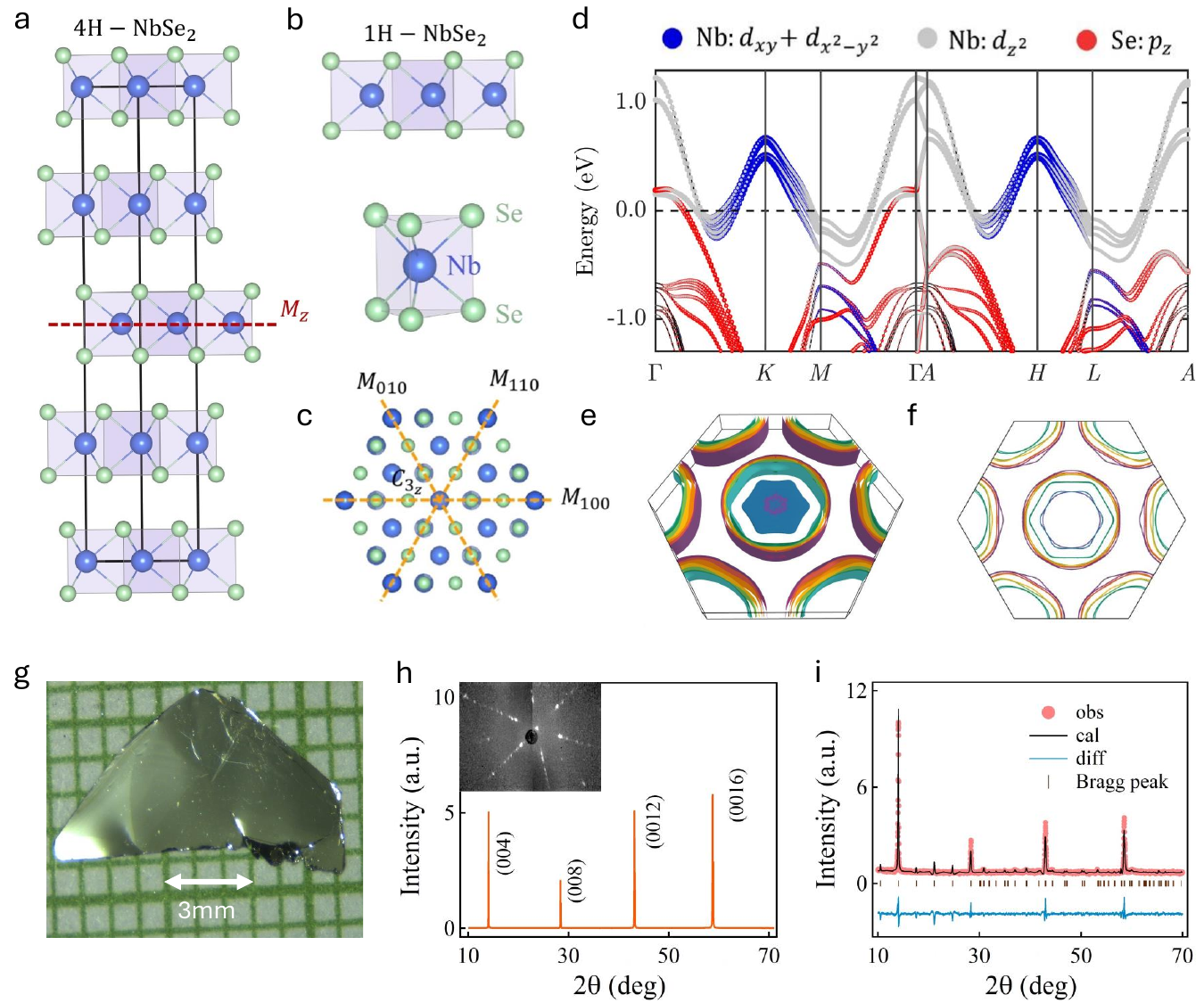}
\caption{\textbf{Crystal lattice and electronic structure of 4$H$-NbSe$_2$.} a, Bulk crystal structure containing four unit cells of monolayer NbSe$_2$ without a center of inversion. The dashed red line marks the $M_z$ mirror plane, which ensures zero $S_x$ and $S_y$ spin components. b, Monolayer NbSe$_2$ without any inversion symmetry and its Nb-Se trigonal local coordination structure. c, Top view of bulk 4$H$-NbSe$_2$, showing hexagonal symmetry with three vertical mirror planes and a three-fold rotational symmetry along $z$-axis. d, Calculated orbital resolved bulk band structure with spin-orbit coupling. The blue, gray and red markers represent Nb $d_{xy}+d_{x^2-y^2}$, Nb $d_{z^2}$, and Se $p_{z}$ orbitals, respectively. e, 3D Fermi surface and corresponding (f) energy contours on the $k_z=0$ plane, revealing multiple 2D Fermi pockets centered at the center and corner of the Brillouin zone. g, Microscopic image of as-grown single crystal of 4$H$-NbSe$_2$. X-ray diffraction patterns: h, for a single crystal with the Laue image in the inset, and i, for polycrystalline 4$H$-NbSe$_2$.} 
\label{XRD} 
\end{figure*}

Among materials exhibiting Ising superconductivity, monolayer 2$H$-NbSe$_2$ stands out due to its noncentrosymmetric crystal structure, strong SOC, and exceptionally high upper critical field ($H_{c2}$) of up to 31.5 T~\cite{ising_nbse2,NbSe2_ising_STM}. Both theoretical and experimental studies have shown that, in addition to Ising SOC, spin-triplet pairing components contribute to enhanced $H_{c2}$ in NbSe$_2$~\cite{ising_with_triplet,mirage_spin_theory_triplet}. As the number of layers increases, the interlayer coupling tends to restore the bulk inversion symmetry in NbSe$_2$, leading to conventional superconducting behavior~\cite{review_ising_sc}. However, recent work has demonstrated the persistence of Ising-like superconductivity in bulk NbSe$_2$ via organic cation intercalation, which preserves Ising SOC without suppressing the superconducting transition temperature~\cite{bulk_nbse2_isng}.\\

Another notable phase, 4$H$-NbSe$_2$, features a noncentrosymmetric crystal structure similar to monolayer 2$H$-NbSe$_2$, but with a doubled unit cell~\cite{NbSe2_rivisit}. This structural inversion symmetry breaking makes 4$H$-NbSe$_2$ a strong candidate to host Ising superconductivity. Unlike monolayer systems, where experimental access is limited, intrinsic bulk single crystals of 4$H$-NbSe$_2$ offer a unique opportunity to investigate the Ising pairing mechanism through bulk-sensitive measurements, providing new insights and a robust platform for exploring Ising superconductivity.

In this work, we report the growth of high-quality single crystals of noncentrosymmetric 4$H$-NbSe$_2$ and provide a comprehensive study of its electronic and superconducting properties. Our experiments confirm a bulk superconducting transition at 6.21(6) K. We observe significant anisotropy in the upper critical field, with the in-plane value exceeding the out-of-plane counterpart by more than six times. Spin-resolved band structure calculations and symmetry analysis further reveal substantial spin splitting and spin-layer locking, supporting the presence of bulk Ising superconductivity in 4$H$-NbSe$_2$.

 \begin{figure*}[tp]
\includegraphics[width=2.0\columnwidth]{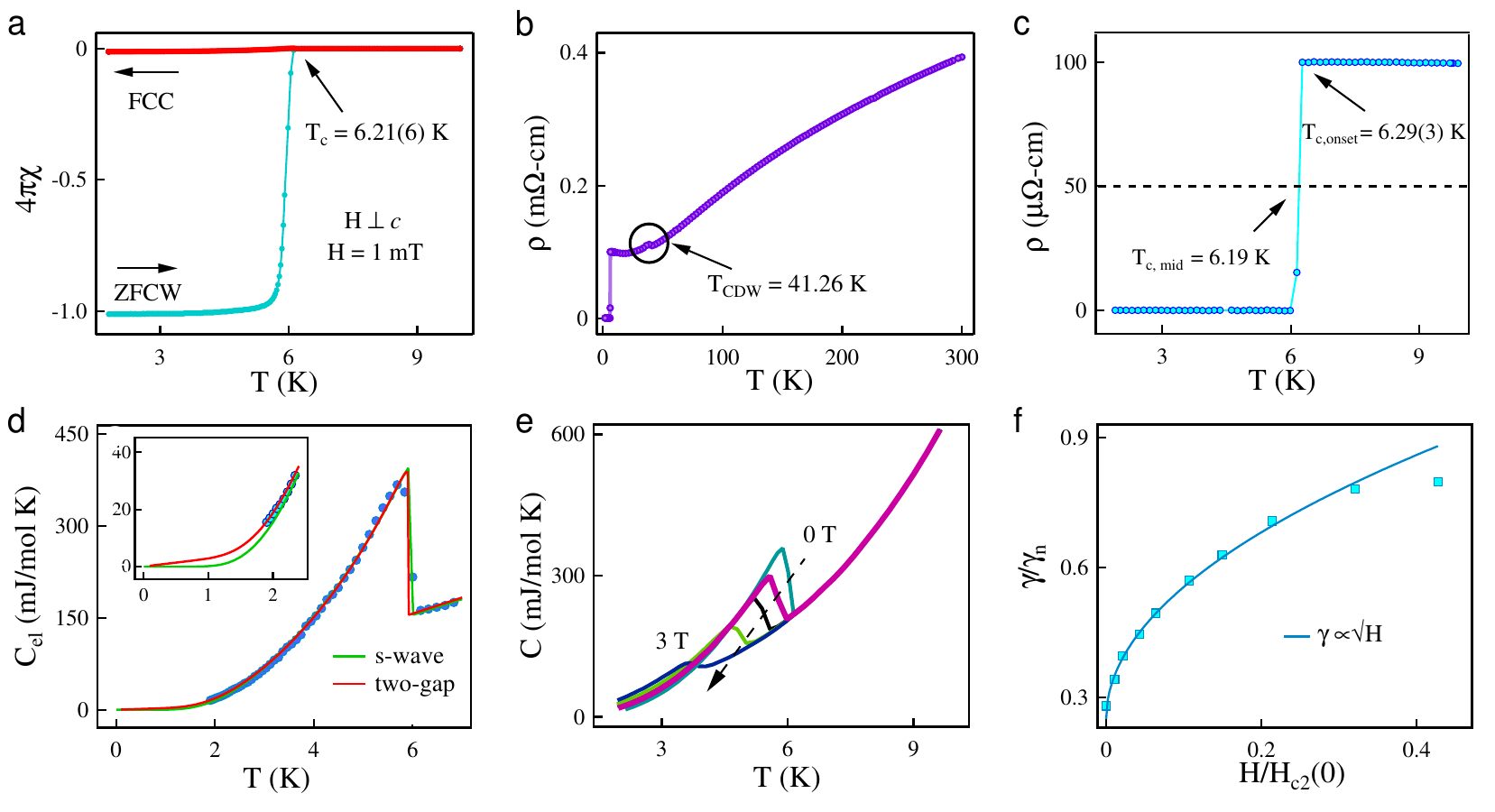}
\caption{\textbf{Magnetic susceptibility and specific heat measurements.} a, Temperature variation of magnetic susceptibility of 4$H$-NbSe$_2$, revealing the superconducting transition at $T_c \sim$ 6.21(6) K for $H \perp c$ orientation. b, Zero field temperature dependence of resistivity up to 300 K, exhibiting CDW transition at 41.26 K. c, Superconducting zero drop in resistivity at the onset of 6.29(3) K. d, Electronic contribution of zero field specific heat data is fitted with s-wave and two-gap $\alpha$-model. The inset shows the zoomed view of the low-temperature fitting. e,~Field-dependent specific heat curves, showing continuously decreasing jump value. f, Sommerfeld coefficient variation with square root field dependence.}
\label{Fig_2}
\end{figure*}

\section{Results}
\section*{Crystal and band structure}
The crystal structure of bulk 4$H$-NbSe$_2$ (Fig.~\ref{XRD}a) belongs to the $P\bar{6}m2$ space group (No. 187), consisting of four NbSe$_2$ monolayers stacked along the hexagonal $c$-axis without a center of inversion. The structure contains a $M_z$ mirror plane, with a reflection point at the center of the unit cell, and exhibits three vertical mirror planes, similar to the monolayer NbSe$_2$ (see Figs.~\ref{XRD}a-c). The orbital-resolved band structure with SOC is shown in Fig.~\ref{XRD}d. It reveals a metallic character with multiple bands intersecting the Fermi level. The highest valence bands at the $K$ and $K'$ valleys are primarily derived from Nb $d_{xy}$ and $d_{x^2-y^2}$ orbitals, while the bands at the $\Gamma$ point are dominated by the Nb $d_{z^2}$ states. These bands form open Fermi sheets, as shown in Figs.~\ref{XRD}e-f, indicating a two-dimensional (2D) character of the Fermi surface. The corresponding Fermi band contours at $k_z = 0$ are illustrated in Fig.~\ref{XRD}f. At the $K$ and $K'$ valleys, each NbSe$_2$ monolayer contributes two Fermi sheets, resulting in a total of eight sheets. Due to the absence of inversion symmetry, these bands are spin-polarized, as discussed below.
 
 Figure~\ref{XRD}g presents a microscopic image of the as-grown single crystal of bulk 4$H$-NbSe$_2$, synthesized using the chemical vapor transport (CVT) method (see experimental details). 
 The X-ray diffraction (XRD) pattern in Fig. \ref{XRD}h reveals a (00$l$) preferred orientation of the grown crystal. The inset further shows the Laue diffraction image, which confirms the hexagonal symmetry of the crystals. The refined XRD pattern of polycrystalline 4$H$-NbSe$_2$ (\figref{XRD}i) confirms its $P\Bar{6}m2$ (187) lattice symmetry with deduced lattice parameters, $a$ = $b$ = 3.44(4) \text{\AA} and $c$ = 25.24(3) \text{\AA}, where $c$-parameter for 4$H$-phase is approximately twice that of the 2$H$ phase.

 \begin{figure*}[tp]
\includegraphics[width=2.0\columnwidth]{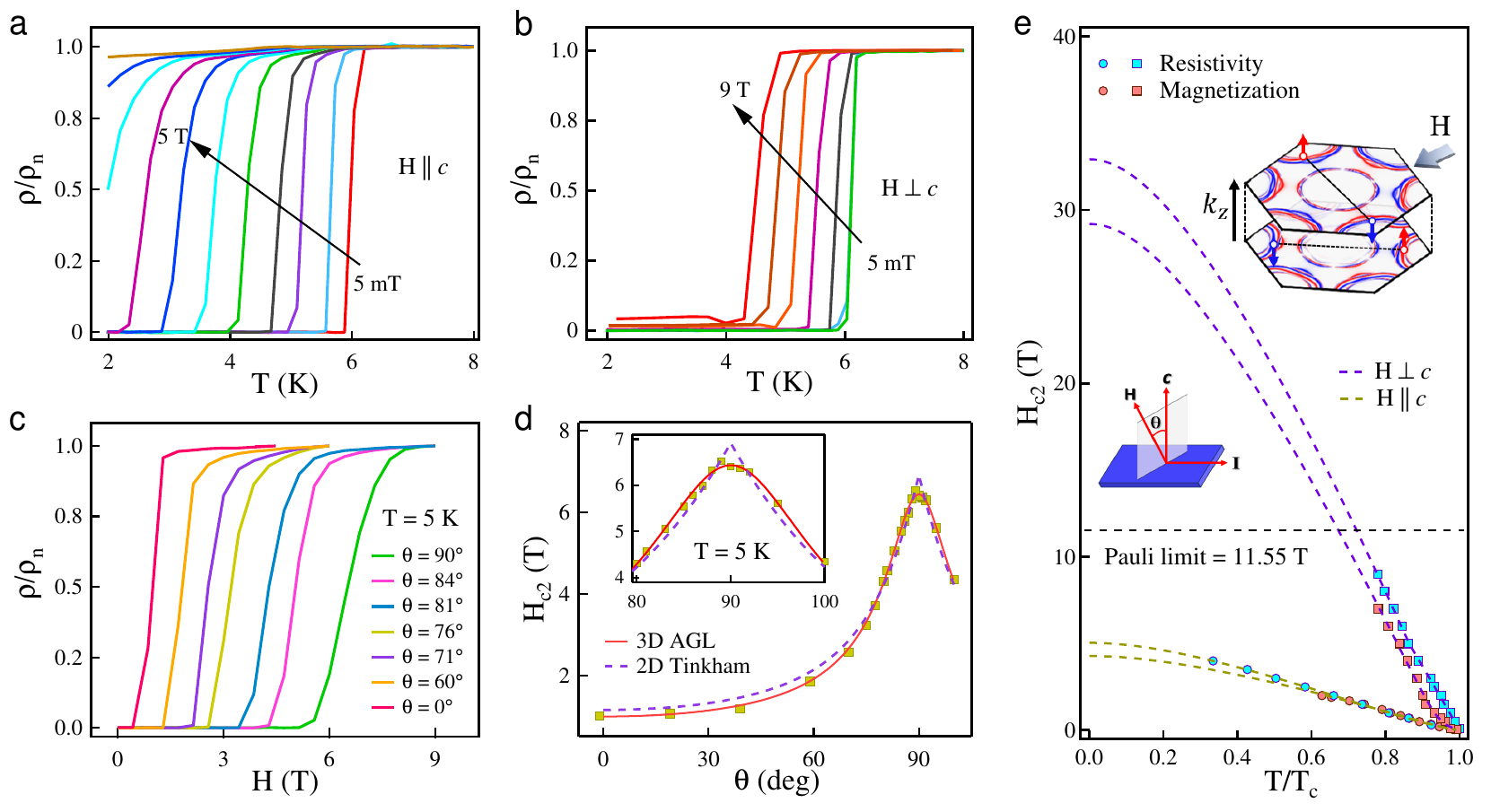}
\caption{\textbf{Resistivity and high upper critical field.} Resistivity variation with temperature for a, $H \parallel c$, and b, $H \perp c$ orientations. c, The anisotropy in the superconducting state for 4$H$-NbSe$_2$ is evident from field-dependent resistivity variations at different $\theta$ values, where $\theta$ is the angle between the magnetic field and the normal to the sample plane. d, $H_{c2}(\theta)$ variation at 5 K is fitted with a 3D anisotropic GL model and 2D Tinkham model, and the inset represents the zoomed part across 90 $\degree$. e, The two-band fitting of the $H_{c2}$ vs $T$ behavior for perpendicular and parallel directions. The black dash line shows the Pauli limit, $H_{c2}^P$ = 1.86$T_c$ = 11.55 T. }
\label{Hc2} 
\end{figure*}

\section*{Superconducting properties}
A comprehensive study of superconductivity in 4$H$-NbSe$_2$ was conducted through low temperature magnetization, resistivity, and specific heat measurements. Magnetization measurement confirmed the superconducting transition temperature ($T_c$) at 6.21(6) K with a clear diamagnetic signal as shown in \figref{Fig_2}a~\cite{poly_4H_NbSe2}. This transition was further confirmed by a sharp drop to zero resistivity, $\rho(T)$, in zero-field at $T^{onset}_c$ = 6.29(3) K (\figref{Fig_2}c). The full range $\rho(T)$ up to 300 K exhibited the charge density wave (CDW) transition for 4$H$-NbSe$_2$ at approximately 41 K, as shown in \figref{Fig_2}b. The temperature variation of the Hall coefficient also confirms the presence of the CDW state (see supplementary Fig. S4b).

Further, zero-field specific heat measurement was performed with the signature of bulk superconductivity at 6.01(1) K. \figref{Fig_2}d shows the electronic contribution ($C_{el}(T)$) of specific heat in the superconducting state, estimated by subtracting the phononic term $\beta_3T^3$ from the total specific heat value (see supplementary for more details). To study the superconducting gap symmetry, $C_{el}(T)$ was fitted by two models: (i) the single gap isotropic $s$-wave model \cite{sp_s_wave_model,sp_alpha_model}, and (ii) the phenomenological two-gap $\alpha$-model \cite{two_gap_specific_heat,sp_nbse2,Mgb2_2_gap_sp_app} (model details in the supplementary). As indicated by \figref{Fig_2}d, the data align well with the two-gap model, particularly at low temperatures (below 2.4 K), suggesting multiband superconductivity in 4$H$-NbSe$_2$ with gap values of 1.75 meV and 1.05 meV, with a fraction of 0.05. However, low-temperature data, extending to $T_c/10$, are needed for a more precise analysis of the superconducting gap symmetry in 4$H$-NbSe$_2$.

The multigap nature of superconductivity in 4$H$-NbSe$_2$ is further corroborated by the field dependence of the Sommerfeld coefficient $\gamma(H)$ below $T_c$ \cite{Yni2b2c}. This coefficient is determined from temperature-dependent specific heat measurements under different applied magnetic fields, shown in \figref{Fig_2}e. In single-gap superconductors, $\gamma(H)$ varies linearly with the magnetic field ($\gamma \propto H$). In contrast, for multi-gap superconductors, the quasiparticle wave functions associated with zero-energy states extend considerably beyond the vortex core, resulting in a distinct behavior where $\gamma \propto \sqrt{H}$ \cite{sp_nbse2}. The calculation of the Sommerfeld coefficient is performed using 
\begin{equation}\label{eqn1}
\frac{C_{\text{el}}}{T} = \gamma + \frac{a}{T} \exp\left(-b\frac{T_c}{T}\right).
\end{equation}
In 4$H$-NbSe$_2$, the Sommerfeld coefficient exhibits square root variation, as depicted in \figref{Fig_2}f. This distinct behavior strongly supports that 4$H$-NbSe$_2$ is likely to be a multi-gap superconductor~\cite{sp_nbse2}.

To estimate the upper critical field $H_{c2}(0)$, we measured the temperature-dependent resistivity under various magnetic fields for two orientations: with the field applied parallel to the $c$-axis ($H \parallel c$), and perpendicular to the $c$-axis ($H \perp c$), as shown in Figs.~\ref{Hc2}a,b. $H_{c2}(T)$ values were estimated at temperatures corresponding to 0.9$\rho_n$ for both orientations. These values were also calculated through magnetization (see Fig.~S2). Observed behavior is inconsistent with the single-gap Ginzburg-Landau (GL) and the Werthamer-Helfand-Hohenberg (WHH) models \cite{hc2_mgb2_fit, whh_model}. Similar behavior has been observed in multi-gap superconductors such as MgB$_2$~\cite{MgB2_Al_doped,mgb2_clean_dirty} and certain iron-based materials~\cite{fe_based_2_gap}. To quantitatively describe this feature, we employed the two-band model~ \cite{lafeos_2_gap} (see Supplementary Materials (SM)). From the fit, the obtained $H_{c2}(0)$ values are $H_{c2}^{\perp c}(0)$ = 32.9(5) T and $H_{c2}^{\parallel c}(0)$ = 5.0(4) T, with the anisotropy factor $(\gamma_{H_{c2}} = H_{c2}^{\perp c}(0)/H_{c2}^{\parallel c}(0)$ = 6.6 for 4$H$-NbSe$_2$. 

\begin{figure*}
\includegraphics[width=1.35\columnwidth]{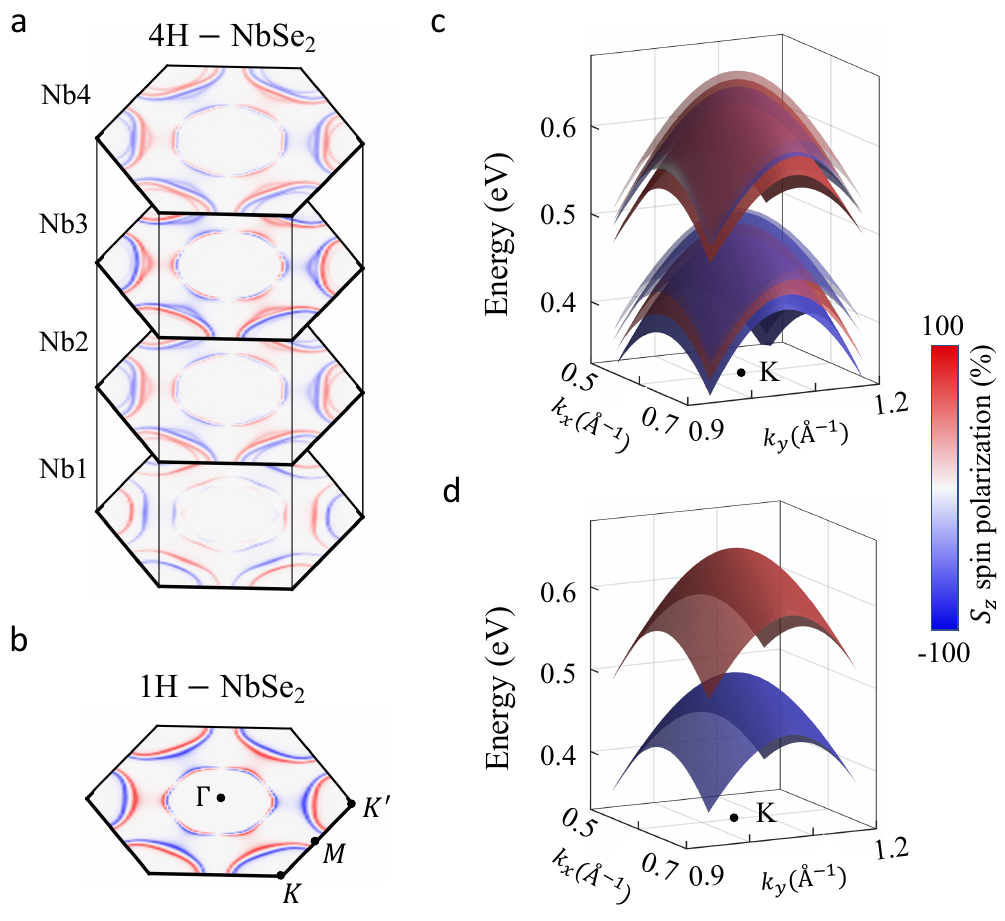}
\caption{\textbf{Spin-polarized states of bulk 4$H$-NbSe$_2$.} (a) Out-of-plane spin-polarized Fermi band contours of bulk 4$H$-NbSe$_2$ at $k_z=0$ plane. The stacking represents layer (or individual Nb atom) resolved spin contribution. (b) The out-of-plane spin-polarized Fermi surface of monolayer 1$H$-NbSe$_2$. $K$ and $K'$ valleys form a time-reversal pair with oppositely oriented spin states. The out-of-plane spin polarization and spin splitting vanish along $\Gamma-M$ mirror lines. (c)-(d) $E-k_x-k_y$ dispersion near the $K$ valley in (c) 4$H$-NbSe$_2$  and (d) 1$H$-NbSe$_2$. The color represents the out-of-plane spin polarization. Each layer in 4$H$-NbSe$_2$ exhibits Zeeman-type spin splitting at the $K$ and $K'$ valleys, akin to that observed in monolayer NbSe$_2$.}
\label{spin} 
\end{figure*}

These results indicate that the in-plane $H_{c2}(0)$ value for 4$H$-NbSe$_2$ is remarkably high relative to other bulk TMD systems~\cite{ising_sc_review,TaS2_ising,NbSe2_rivisit}. In type-II superconductors, Cooper pairs break via orbital and Pauli paramagnetic effects~\cite{orbital_paramagnetic_limit}. In the orbital effect, pairs break due to increased kinetic energy, with the orbital limiting field ($H_{c2}^{orbital}$) determined using the WHH expression. The initial slope of temperature-dependent $H_{c2}$ gives $H_{c2}^{orbital}$ = 22.36(7) T. This high value suggests the presence of multiband effects, strong 3D orbital effects, or decoupling layers at low temperatures~\cite{orbital_limit_effect}. On the other hand, the Pauli paramagnetic effect appears when electron spins align with the magnetic field, breaking Cooper pairs. The Pauli paramagnetic limit, $H_{c2}^{P}$ is expressed as 1.86 $T_c$, yields 11.55(1) T~ \cite{graphene_symmetry}. A violation of this limit originates from strong spin-orbit coupling, spin-triplet pairing, or the occurrence of the Fulde-Ferrell-Larkin-Ovchinnikov (FFLO) state~\cite{theory_ising_sc,Ba6Nb11S28,bulk_pauli,graphene_symmetry,Ba6Nb11Se28_ising}. For 4$H$-NbSe$_2$, the obtained in-plane $H_{c2}(0)$ value is nearly three times higher than the Pauli limit. A similar high value has been observed in the bilayer, trilayer, and organic cation-intercalated 2$H$-NbSe$_2$~\cite{mirage_spin_theory_triplet,bulk_nbse2_isng}. Notably, in the monolayer case, the upper critical field exceeds the Pauli limit by a factor of six, which has been attributed to Ising superconductivity or the presence of spin-triplet pairing, supported by the observation of a 76 meV spin-splitting energy \cite{ising_nbse2,ising_with_triplet}. For our 4$H$-NbSe$_2$ sample, the high $H_{c2}$ aligns with the band structure calculations, revealing a spin-splitting energy of $\sim$ 150 meV, strongly suggesting the Ising pairing mechanism in the bulk form. Furthermore, angle-resolved photoemission spectroscopy (ARPES) study of intercalated 2$H$-NbSe$_2$, showing dispersed bands similar to monolayer 1$H$-NbSe$_2$, supporting the Ising-like interaction as a key factor in Pauli limit violation~\cite{bulk_nbse2_isng}. Additionally, the Maki parameter $\alpha_m$ = $\sqrt{2}\frac{H_{orb}(0)}{H_P(0)}$ is calculated for 4$H$-NbSe$_2$ and found to be approximately 2.7, which is significantly large. A large Maki parameter ($\alpha_m > 1.7$) suggests the possible emergence of a FFLO state, as previously observed in 2$H$-NbSe$_2$ \cite{NbSe2_FFLO}.

The anisotropic behavior of superconductivity in 4$H$-NbSe$_2$ was investigated by field-dependent resistivity measurements at various angles $\theta$, as illustrated in \figref{Hc2}d. The obtained $H_{c2}(\theta)$ variation (\figref{Hc2}e), measured at $T$ = 5 K was explained by the two models: the 3D anisotropic Ginzburg-Landau (AGL) model and the 2D Tinkham model \cite{AuSn4} (see SMs for model details). The zoomed part of fitting across 90$\degree$ in the inset of \figref{Hc2}d demonstrates better fitting with the 3D AGL model. The fitting parameters are $H_{c2}^{\parallel}$($0\degree$) = 0.98(2) T, and $H_{c2}^{\perp}$($90\degree$) = 6.43(4) T at 5 K with an anisotropy factor of 6.56 (nearly equal to the anisotropy factor observed from multi-gap model fitting \figref{Hc2}c). This 3D superconducting behavior with high anisotropy factor was reported in bulk 2$H$-NbSe$_2$~\cite{NbSe2_angle}.\\


\section*{Spin-polarized states and spin splitting} Figure~\ref{spin} presents the spin-polarized states of bulk 4$H$-NbSe$_2$  and monolayer 1$H$-NbSe$_2$ and their momentum distribution within the Brillouin zone. Due to the absence of inversion symmetry and the presence of the $M_z$ mirror plane, all states away from time-reversal invariant points exhibit spin splitting with out-of-plane spin polarization in bulk 4$H$-NbSe$_2$. This spin splitting is most pronounced in the $K$ and $K'$ valleys, which form Kramers doublets, each featuring four pairs of spin-split bands (Fig.~\ref{spin}a). Each NbSe$_2$ layer contributes one pair of spin-polarized states, as illustrated in the layer-resolved band structure (Fig.~\ref{spin}c). Within a given NbSe$_2$ layer, the spin splitting remains nearly constant at approximately 150 meV. In contrast, the spin splitting between states arising from different layers varies depending on their relative positions in the unit cell. Adjacent layers exhibit greater splitting due to enhanced interlayer coupling. Despite the 2D nature of the Fermi surface, the out-of-plane coupling between layers remains finite. Figure~\ref{spin}b presents the spin-polarized constant energy contours and energy dispersion for monolayer 1$H$-NbSe$_2$, highlighting valley-selective spin splitting and out-of-plane spin polarization as found in the bulk.

Finally, we discuss the constraints of our findings for the possible superconducting order parameters. Both the large anisotropy of the critical field seen in experiments and the typical spin-orbit splitting of the Fermi surfaces in our first-principles calculations being much larger than the superconducting energy scales imply that we can use the weak-pairing description of \refcite{WeakPairingDescr}. In this limit, the superconducting order parameter is just a complex function $\Delta_{\vec{k}a}$ of the momentum $\vec{k}$ on each Fermi sheet $a$; for a superconducting order parameter transforming under irreducible representation (IR) $n$, $\Delta_{\vec{k}a}$ will be a linear superposition $\Delta_{\vec{k}a} = \sum_{\mu}\eta_\mu \varphi^\mu_{\vec{k}a}$, $\eta_\mu \in \mathbb{C}$, of the basis functions $\varphi^\mu_{\vec{k}a}$, of $n$, which can be chosen to be real. Since the superconducting gap is given by $|\Delta_{\vec{k}a}|$, we can exclude all IRs ($A_1''$, $A_2''$, and $E''$) of the point group $D_{3h}$ which are odd under $M_z$: Fermi surfaces in \figref{XRD}e would then give rise to nodal lines in the $k_z=0,\pi$ planes, which is not consistent with our specific heat data. Similarly, any state transforming under IR $A_2'$ will be odd under the mirror symmetries in \figref{XRD}c and thus exhibit nodal lines. We are then left with only two possibilities: a state transforming under the trivial IR $A_{1}'$ or a more exotic time-reversal-symmetry-breaking chiral superconductor transforming under $E'$, both of which are fully gapped. We further note that there are no symmetries relating the pockets around $K^{(\prime)}$ and $\Gamma$, such that the gap of both of these two candidates will also be generically different on those pockets, in line with our experimental findings. If the pairing mechanism is primarily electron-phonon coupling, only the $A_{1}'$ state is possible \cite{PairingMechanism}; furthermore, in this scenario, the sign of (now real) $\Delta_{\vec{k}a}$ will be the same on spin-orbit-split Fermi surfaces. As Ising spin-orbit coupling seems to dominate, this will translate into a superconductor with the spin-singlet component being larger than the (primarily out-of-plane, unitary) triplet part. If, instead, fluctuations of a time-reversal-odd collective mode induce pairing---something that is not implausible since NbSe$_2$ is believed \cite{CloseToMagneticInstab1,CloseToMagneticInstab2} to be close to a magnetic instability---both states can be stabilized. As the chiral state would be more sensitive to disorder, it is a generically less natural candidate state, but future studies are required to determine whether it is realized or not. To this end, $\mu$SR or disorder-dependent studies would be well suited.

\section{Conclusion}
In summary, we have grown single crystals of a new inversion-asymmetric phase of transition metal dichalcogenides, 4$H$-NbSe$_2$, and investigated its superconducting properties. Superconductivity is observed to emerge at 6.21(6) K. Evidence for possible multi-gap superconductivity, inferred from specific heat measurements, upper critical fields, and the breaking of the Pauli limit, points to an unconventional nature of this system. The angular dependence of the upper critical field reveals a three-dimensional superconducting behavior. Additionally, the in-plane upper critical field exceeds the out-of-plane counterpart by more than a factor of six and is nearly three times greater than the Pauli limiting field, as well as the spin-splitting energy of approximately 150 meV, similar to the behavior observed in the bilayer and trilayer 2$H$-NbSe$_2$. These results confirms the presence of Ising superconductivity in 4$H$-NbSe$_2$. Our findings enhance the understanding of interlayer coupling in Ising superconductivity, providing insights that could lead to the discovery of new Ising superconductors. The bulk nature of the sample allows for further investigations into the superconducting ground state, which is not feasible in atomically thin counterparts. Thus, the noncentrosymmetric bulk 4$H$-NbSe$_2$ serves as an ideal platform to explore pairing mechanisms in Ising superconductivity and its potential applications in quantum devices.

\section*{Acknowledgments}

R.P.S. acknowledges the Science and Engineering Research Board, Government of India, for the Core Research Grant No. CRG/2023/000817. The work at TIFR Mumbai is supported by the Department of Atomic Energy of the Government of India under Project No. 12-R\&D-TFR-5.10- 0100 and benefited from the HPC resources of TIFR Mumbai.


All data that support the findings of this study are available within the paper and/or Supplemental Materials. Additional data related to this paper may be requested from R.P.S. (rpsingh@iiserb.ac.in).

\end{document}